\documentclass{llncs}
\usepackage{makeidx}
\begin{document}

\newcommand{\msum}{\mathop{\sum}\limits}
\newcommand{\rrang}{{\rm{rang}}\;}
\newcommand{\rmax}{\mathop{\rm max}\limits}
\newcommand{\rmin}{\mathop{\rm min}\limits}

\title{Affine Transformations of Loop Nests\\
       for Parallel Execution\\
       and Distribution of Data over Processors
       \thanks{Supported by the Belorussian Foundation for Basic
               Research under Grant F05--019}}
\author{E.V.~Adutskevich, S.V.~Bakhanovich, \and N.A.~Likhoded}
\institute{National Academy of Sciences of Belarus,\\
     Institute of Mathematics, \\
     Surganov str., 11 , Minsk 220072 BELARUS \\
     Phone: 375 17 284 26 43  \ \ Fax: 375 17 284 09 15  \\
         \email{\{zhenya, bsv, likhoded\}@im.bas-net.by}}
\maketitle

\begin{abstract}
The paper is devoted to the problem of mapping affine loop nests
onto distributed memory parallel computers. A method to find
affine transformations of loop nests for parallel execution and
distribution of data over processors is presented. The method
tends to minimize the number of communications between processors
and to improve locality of data within one processor. A problem of
determination of data exchange sequence is investigated.
Conditions to determine the ability to arrange broadcast is
presented.

\end{abstract}

\section{Introduction}

To map algorithms given by sequential programs onto distributed
memory parallel computers is to distribute data and computations
to processors, to determine an execution sequence of operations
and a data exchange sequence. The most important problems are:
scheduling \cite{voev}, space-time mapping \cite{lim:lam},
alignment \cite{voev,di:rob,likh}, determination of data exchange
sequence \cite{di:rand:rob}. An essential stage of the solution of
these problems is to find functions (scheduling functions,
statement and array allocation functions) satisfying certain
constraints.

One of the preferable parallelization schemes is based on
obtaining multi-dimensional scheduling functions. Some coordinates
of the multi-dimensional scheduling functions are used for
operations allocation. The other coordinates are used for
scheduling operations.

For the program execution time to be as small as possible it is
necessary to solve an alignment problem. It consists in
coordinated operations and data allocation to minimize the
communications.

The program execution time depends not only on the execution time
of operations but also on the memory access time. The access time
depends on data location in the hierarchical memory. Therefore the
problem of prompt data reuse within one processor (localization
problem) is of great importance \cite{ahm:mat:ping}. Two kinds of
localization, such as localization in time and in space, are
available. The time localization is used to set the operations
execution sequence so that the data be reused before it is moved
to a lower level memory. The spatial localization allows to use
the data allocated close to each other in memory. The locality
depends on the execution sequence of operations. Hence, it is
desirable to take it into account when scheduling functions are
obtained.

In this paper, a method of simultaneous solution of all mentioned
problems is suggested. It results in high performance of parallel
code execution.

After transformation of an algorithm for parallel execution it is
necessary to determine the data exchange sequence. In many cases
utilization of broadcast, gather, scatter, reduction, and
translation enables to improve the efficiency of a parallel
program. In this paper, we investigated the problem of
determination of the data exchange sequence. We suggest the
conditions to determine the case when the broadcast communications
may be used.

\section{Main Definitions}

Let an algorithm be represented by an affine loop nest. For such
algorithms, array indices and bounds of loops are affine functions
of outer loop indices or loop-invariant variables. Let a loop nest
contain $K$ statements $S_\beta$ and use $L$ arrays $a_l$. By
$V_\beta$ denote the iteration domain of statement $S_\beta,$ by
$W_l$ denote the index domain of array $a_l.$ By $n_\beta$ denote
a number of loops surrounding statement $S_\beta.$ By $\nu_l$
denote dimension of array $a_l.$ Then
$V_\beta\subset\bbbz^{n_\beta},\ W_l\subset\bbbz^{\nu_l}.$ By
$J\in\bbbz^{n_\beta}$ denote the iteration vector, by
$N\in\bbbz^e$ denote the vector of outer variables, $e$ is the
number of these variables.

Let $\overline F_{l,\beta,q}\colon V_\beta\to W_l$ denote access
function that puts the iteration domain $V_\beta$ into
correspondence with the index domain $W_l$ for the $q$-th input of
elements of array $a_l$ into instruction $S_\beta.$ Suppose
$\overline F_{l,\beta,q}$ are affine functions: $
    \overline F_{l,\beta,q}(J)=F_{l,\beta,q}J + G_{l,\beta,q}N + f^{(l,\beta,q)},
$
where $J\in V_\beta,$ $F_{l,\beta,q}\in\bbbz^{\nu_l\times
n_\beta},$ $N\in\bbbz^e,$ $G_{l,{\beta},q} \in \bbbz^{\nu_l\times
e},$ $f^{(l,{\beta},q)} \in \bbbz^{\nu_l}.$

Given a statement $S_\beta$, a computation instance of $S_\beta$
is called an operation and is denoted by $S_\beta(J).$ Denote a
dependence of operation $S_\beta(J)$ from operation $S_\alpha(I)$
by $S_\alpha(I)\to S_\beta(J).$ We consider flow-, anti-, out-,
and in-dependences. Denote by P a set of pairs of indices such
that $S_\alpha(I)\to S_\beta(J).$

Let $\overline\Phi_{\alpha,\beta} \colon V_{\alpha,\beta} \to
V_\alpha$ be dependence function. If $S_\alpha(I) \to S_\beta(J)$,
$I\in V_\alpha,\ J\in V_{\alpha,\beta} \subseteq V_\beta$, then
$I=\overline\Phi_{\alpha,\beta}(J).$ Suppose
$\overline\Phi_{\alpha,\beta}$ are affine functions: $
        \overline\Phi_{\alpha,\beta}(J) =
        \Phi_{\alpha,\beta}J + \Psi_{\alpha,\beta}N -
        \varphi^{(\alpha,\beta)},\ \
        J \in V_{\alpha,\beta},\
$\ \ $\Phi_{\alpha,\beta}\in\bbbz^{n_\alpha\times n_\beta}$,
$\varphi^{(\alpha,\beta)}\in\bbbz^{n_\alpha}$, $N\in\bbbz^e,$
$\Psi_{\alpha,\beta}\in\bbbz^{n_\alpha\times e}.$

\section{Multi-Dimensional Scheduling Functions. Data Allocation Functions}

Let $n=\rmax_{1\le\beta\le K} n_\beta.$ Let functions $\overline
t^{(\beta)}\colon V_{\beta}\to\bbbz^n$ assign a vector
$(t_1^{(\beta)}(J),\ldots,\linebreak t_n^{(\beta)}(J))$ to each
operation $S_\beta(J).$ Suppose $t_\xi^{(\beta)}$ are affine
functions:
$        t_\xi^{(\beta)}(J)=\tau^{(\beta,\xi)}J + b^{(\beta,\xi)}N +
            a_{\beta,\xi},\ \ J\in V_{\beta},\
            \tau^{(\beta,\xi)}\in\bbbz^{n_\beta},\
            b^{(\beta,\xi)},N\in\bbbz^e,\ a_{\beta,\xi}\in\bbbz,\
            1\le\beta\le K,\ 1\le\xi\le n.
$

Functions $\overline t^{(\beta)}$ are called vector scheduling
functions if
\begin{eqnarray}
    &\rrang T^{(\beta)} = n_\beta,\ 1\le\beta\le K \enspace,&
\label{rang}\\
    &\overline t^{(\beta)}(J) \ge_{lex} \overline t^{(\alpha)}(I),\quad
        J\in V_\beta,\ I\in V_\alpha,\ \ {\rm if}\ S_\alpha(I) \to S_\beta(J)
        \enspace.&
\label{lexge}
\end{eqnarray}
Here $T^{(\beta)}$ is a matrix whose rows are vectors
$\tau^{(\beta,1)},\ldots,\tau^{(\beta,n)},$ $S_\alpha(I) \to
S_\beta(J)$ is any dependence except in-dependence, notation
$\ge_{lex}$ denotes "lexicographically greater or equal to".

A set of vector functions $\overline t^{(\beta)},\ 1\le\beta\le
K,$ is called a multi-dimensional scheduling. We can use these
functions to transform loops assuming the operation $S_{\beta}(J)$
to be executed at the iteration $\overline t^{(\beta)}(J).$ Thus
we interpret elements of the vector $\overline t^{(\beta)}$ as
indices of the transformed loop nest for the statement $S_\beta:$\
\ $t_1^{(\beta)}$ is the index of the outermost loop,
$t_n^{(\beta)}$ is the index of the innermost loop. Note that the
functions $\overline t^{(\beta)}$ determine permissible
transformation of the loop nest, i.e., this transformation keeps
the execution sequence of dependent operations.

We consider functions $(t^{(\beta)}_1,\ldots,t^{(\beta)}_r),\
r<n,$ as allocation functions that determine spatial mapping of an
algorithm to $r$-dimensional space of virtual processors. That is,
the values of indices of $r$ external loops of the transformed
algorithm determine processor coordinates. The values of indices
of $n-r$ internal loops determine iterations to be executed on the
processor.

Usually we need to take into account the number of processors used
to execute the program. Then to simplify code generation  it is
necessary that the following conditions be valid
$t^{(\beta)}_\xi(J)\ge t^{(\alpha)}_\xi(I),\ J\in V_\beta,\ I\in
V_\alpha,\ {\rm if}\ S_\alpha(I) \to S_\beta(J),\ 1\le\xi\le r.$

Let functions $\overline d^{(l)} : W_l\to\bbbz^r$ assign a vector
$(d^{(l)}_1(F),\ldots,d^{(l)}_r(F))$  to each element $a_l(F)$ of
an array $a_l$. Suppose $d^{(l)}_\xi$ are affine functions: $
    d^{(l)}_\xi(F)=\eta^{(l,\xi)}F + z^{(l,\xi)}N + y_{l,\xi},\ \ F\in W_l,\
    \eta^{(l,\xi)}\in \bbbz^{\nu_{l}},\
    z^{(l,\xi)},N\in\bbbz^e,\ y_{l,\xi}\in \bbbz,\
    1\le l\le L,\ 1\le\xi\le r.
$ Let element $a_l(F)$ be stored in the local memory of the
processor determined by the coordinates
$(d^{(l)}_1(F),\ldots,d^{(l)}_r(F)).$

Let us introduce some notation:\ $\widetilde\tau^{(\xi)} =
(\tau^{(1,\xi)}, \ldots ,\tau^{(K,\xi)},\eta^{(1,\xi)},\ldots,
    \eta^{(L,\xi)},\linebreak
    b^{(1,\xi)},\ldots,
    b^{(K,\xi)},z^{(1,\xi)},\ldots,z^{(L,\xi)},a_{1,\xi}, \ldots ,
    a_{K,\xi},y_{1,\xi},\ldots,y_{L,\xi})$
    is a vector, whose entries are parameters of functions
    $t_\xi^{(\beta)}$ and $d^{(l)}_\xi.$

The following proposition gives the condition to be used for
finding scheduling functions that satisfy condition~(\ref{rang}).

\begin{proposition}\label{prop}
    Suppose $\rrang T^{(\beta)}_{1:{\xi}-1}=r,\ r<n_\beta,$ where
    $T^{(\beta)}_{1:{\xi}}$ is a matrix whose rows are vectors
    $\tau^{(\beta,i)},\ 1\le i\le\xi.$ Suppose
    $s_\beta^{({\xi})}$ is a fixed vector of a set
    $S_\beta^{({\xi})}=\{\, s\in\bbbz^{n_\beta} \mid \tau^{(\beta,i)}
    s=0,\ 1\le i\le\xi-1,\ s\ne 0 \,\},\ 2\le\xi\le n.$
    Then $\rrang T^{(\beta)}_{1:{\xi}}=r+1$ if
    $\tau^{(\beta,\xi)} s_\beta^{(\xi)} \neq 0.$
\end{proposition}

Condition $\tau^{(\beta,\xi)} s_\beta^{(\xi)} \neq 0$ is
equivalent to the following inequality in the vector form
    \begin{equation}
        \big|\widetilde\tau^{(\xi)} \widetilde s_\beta^{({\xi})}\big| \ge 1\enspace.
    \label{trang}
    \end{equation}

Let $v^{(\alpha,\beta,m)}$ be vertices of the polyhedron
$V_{\alpha,\beta},$ $m(\alpha,\beta)$ be the number of the
vertices. Any vertex $v^{(\alpha,\beta,m)}$ can be represented in
the form $v^{(\alpha,\beta,m)}= R^{(\alpha,\beta,m)} N +
\omega^{(\alpha,\beta,m)}.$ Let $N^{(0)}\in\bbbz^e$ be a vector
whose $i$-th entry is equal to the smallest possible value of the
outer variable $N_i.$ Suppose coordinates of the vector $N$ can be
unlimited large. Then we can show that
$t_{\xi}^{(\beta)}(J)-t_{\xi}^{(\alpha)}(I)$ is non-negative for
all $I$ and $J$ such that $S_{\alpha}(I)\to S_{\beta}(J)$  iff
\vspace{-2mm}
$$
    \begin{array}{c}
        \big((\tau^{(\beta,\xi)} - \tau^{(\alpha,\xi)}
        \Phi_{\alpha,\beta}) R^{(\alpha,\beta,m)}+b^{(\beta,\xi)} -
        b^{(\alpha,\xi)} -
        \tau^{(\alpha,\xi)}\Psi_{\alpha,\beta}\big)N^{(0)}+\\
        +(\tau^{(\beta,\xi)}-\tau^{(\alpha,\xi)} \Phi_{\alpha,\beta})
        \omega^{(\alpha,\beta,m)}+a_{\beta,\xi} - a_{\alpha,\xi} +
        \tau^{(\alpha,\xi)} \varphi^{(\alpha ,\beta)}\ge 0,\
        1\le m\le m(\alpha,\beta);
    \end{array}
$$
$$
    (\tau^{(\beta,\xi)} - \tau^{(\alpha,\xi)} \Phi_{\alpha,\beta})
    R^{(\alpha,\beta,m)}+b^{(\beta,\xi)} - b^{(\alpha,\xi)}
    -\tau^{(\alpha,\xi)}\Psi_{\alpha,\beta}\ge 0,\
    1\le m\le m(\alpha,\beta).
$$
\vspace{-2mm}
In the vector-matrix form
\begin{equation}
        \widetilde{\tau}^{({\xi})} D_{\alpha, \beta}^{\varphi} \ge 0,\quad
        \widetilde{\tau}^{({\xi})} D_{\alpha, \beta} \ge 0 \enspace.
        \label{depend}
\end{equation}

Let introduce in the consideration vector variables
$z_{\alpha,\beta}^{\varphi}$ and $z_{\alpha,\beta}.$ The solution
of~(\ref{depend}) is the solution of equations
\begin{equation}
    \widetilde{\tau}^{({\xi})}D_{\alpha,\beta}^{\varphi}-
        z_{\alpha,\beta}^{\varphi}=0,\ \
        z_{\alpha,\beta}^{\varphi}\ge 0, \quad
    \widetilde{\tau}^{({\xi})}D_{\alpha,\beta}-
        z_{\alpha,\beta}=0,\ \
        z_{\alpha,\beta}\ge 0 \enspace.
\label{depend_z}
\end{equation}

The following propositions can be easily proved:

1) If $z_{\alpha,\beta}^{\varphi}=0,$ $z_{\alpha,\beta}=0$
in~(\ref{depend_z}) for all $\xi,\ 1\le\xi\le n,$
then $\overline{t}^{(\beta)}(J)=\overline{t}^{(\alpha)}(I)$ for
all $I$ and $J$ such that $S_{\alpha}(I)\to S_{\beta}(J).$

2) If $z_{\alpha,\beta}^{\varphi}>0$ in~(\ref{depend_z}), then
$t_{\xi}^{(\beta)}(J)-t_{\xi}^{(\alpha)}(I)>0$ for all $I$ and $J$
such that $S_{\alpha}(I)\to S_{\beta}(J).$

Thus, to find space-time mapping of an algorithm is to find
vectors $\widetilde\tau^{(\xi)},\ 1\le\xi\le n,$ which the
following conditions are valid for.
Suppose we are searching vectors $\widetilde\tau^{(\xi)},\
\xi=1,2,\ldots,n,$ sequentially. Then condition~(\ref{trang}) has
to be valid for $\beta$ such that $n-\xi+1=n_{\beta}-\rrang
T^{(\beta)}_{1:{\xi-1}}.$
Conditions~(\ref{depend_z}) have to be valid for all
$(\alpha,\beta)\in P$ except the following case.
Suppose $\xi\ge r+1$ and for some $(\alpha,\beta)\in P$ the
inequality $z_{\alpha,\beta}^\varphi>0$ is valid, then validity of
conditions~(\ref{depend_z}) is not necessary for these
$(\alpha,\beta)$ in the sequel.

Consider the alignment problem. The operation $S_\beta(J)$ is
assigned to execute at the virtual processor
$(t^{(\beta)}_1(J),\ldots,t^{(\beta)}_r(J)).$ The array element
$a_l(\overline F_{l,\beta,q}(J))$ is stored in the local memory of
the processor $(d^{(l)}_1(\overline{F}_{l,\beta,q}(J)), \ldots,
d^{(l)}_r(\overline{F}_{l,\beta,q}(J))).$ The expressions
$\delta^{l,\beta,q}_\xi (J)= t^{(\beta)}_\xi(J) -
d^{(l)}_\xi(\overline{F}_{l,\beta,q} (J)),\ 1\le\xi\le r,$
determine the distance between the processors. Assuming
$\delta^{l,\beta,q}_\xi (J)=0$ we obtain conditions for
communication-free allocation:
    $\tau^{(\beta,\xi)} -\eta^{(l,\xi)} F_{l,\beta,q}=0,$
    $b^{(\beta,\xi)} - \eta^{(l,\xi)} G_{l,\beta,q} - z^{(l,\xi)}=0,$
    $a_{\beta,\xi} - \eta^{(l,\xi)} f^{(l,\beta,q)} - y_{l,\xi}=0.$
In the vector-matrix form
$
    \widetilde\tau^{(\xi)} \Delta^F_{l,\beta,q} = 0,\ 
    \widetilde\tau^{(\xi)} \Delta^G_{l,\beta,q}=0,\ 
    \widetilde\tau^{(\xi)} \Delta^f_{l,\beta,q}=0. 
$

Introduce in the consideration vector variables $z^F_{l,\beta,q},$
$z^G_{l,\beta,q},$  $z^f_{l,\beta,q}.$ Thus, to find operation and
data allocation such that a number of communications is as small
as possible is to minimize (or to put to zero if it is possible)
coordinates of the vectors $z^F_{l,\beta,q},$ $z^G_{l,\beta,q},$
$z^f_{l,\beta,q}$ which the following equations are valid for
\begin{equation}
    \big|\widetilde\tau^{(\xi)} \Delta^F_{l,\beta,q}\big| - z^F_{l,\beta,q} = 0,\
    \big|\widetilde\tau^{(\xi)} \Delta^G_{l,\beta,q}\big| - z^G_{l,\beta,q}=0,\
    \big|\widetilde\tau^{(\xi)} \Delta^f_{l,\beta,q}\big| - z^f_{l,\beta,q}=0
    \enspace.
\label{Joff_z}
\end{equation}
Here $|v|$ is a vector whose entries are modules of entries of a
vector $v.$

\section{Conditions of Time and Space Localization}

To obtain time localization is to find functions
$\overline{t}^{(\beta)}$ so that values
$\overline{t}^{(\beta)}(J)$ and $\overline{t}^{(\beta)}(I)$
satisfying~(\ref{lexge}) are as more lexicographically close to
each other as it is possible (reuse of an array element is as more
quicker as these values are closer). We reduced
conditions~(\ref{lexge}) to constraints~(\ref{depend_z}); thus, to
achieve our goal is to minimize (to zero at best) vectors
$z_{\alpha,\beta}^\varphi$ and $z_{\alpha,\beta}.$

Validity of condition~(\ref{lexge}), i.e.,
conditions~(\ref{depend_z}) is necessary for all dependences
except in-dependences. Write analogues of
conditions~(\ref{depend_z}) for in-dependences:
\begin{equation}
    \big| \widetilde{\tau}^{({\xi})}D_{\alpha,\beta}^{\varphi}\big|
        - z_{\alpha,\beta}^{\varphi} =0,\quad
    \big|\widetilde{\tau}^{({\xi})}D_{\alpha,\beta}\big|
        - z_{\alpha,\beta}           =0 \enspace,
\label{depend_z_in}
\end{equation}
Thus, requirements of time localization can be reduced to vectors
$z_{\alpha,\beta}^\varphi$ and $z_{\alpha,\beta}$ minimizing
(zeroing if it is possible) when conditions~(\ref{depend_z}),
(\ref{depend_z_in}) are valid.

To obtain space localization is to use array elements that are
stored close to each other in memory at the iterations that are
close to each other. To be definite, assume that we use a
programming language C. In this case, storing array elements is
realized by rows. Thus, the $l$-th array elements that are stored
close to each other in memory are those that differ from each
other in the last coordinate of the index expressions: $
    \overline{F}_{l,\beta ,q}(J)-\overline{F}_{l,\beta ,q}(I)=
    \lambda e^{(\nu_l)}_{\nu_l},\ \lambda\in\bbbz.
$ We realize space localization among operations of the same
statement for the fixed access to array.

Introduce some notation: $\widetilde{F}_{l,\beta
,q}\in\bbbz^{(\nu_l-1)\times n_\beta}$  is a matrix whose rows are
rows of the matrix $F_{l,\beta ,q}$ except the last row;\
$r(l,\beta,q)$ is rang of the matrix $\widetilde{F}_{l,\beta ,q},$
$r(l,\beta,q)<n_\beta;$\ $d^{(\gamma)}_{{l,\beta ,q}}=0,\
1\le\gamma\le n_\beta-r(l,\beta,q)),$ is a fundamental system of
solutions of a uniform system of equations $\widetilde{F}_{l,\beta
,q}x=0.$

\begin{theorem}
    Let $\overline t^{(\beta)}$ be a multi-dimensional scheduling.
    Choose functions $t_{\xi}^{(\beta)},$
    $\xi\in\{\xi_1,\ldots,\xi_{r(l,\beta,q)}\},$ among functions
    $t_1^{(\beta)},\ldots,t_d^{(\beta)},\ d<n.$ Suppose these
    functions satisfy conditions
    \begin{eqnarray}
        &\tau^{(\beta,\xi)}d^{(\gamma)}_{{l,\beta ,q}}=0 \enspace,\quad
        1\le\gamma\le n_\beta-r(l,\beta,q)\enspace,
        &\label{splcon1}\\
        &\rrang T_{l,\beta,q}=r(l,\beta,q) \enspace.
        & \nonumber 
    \end{eqnarray}
    Here $T_{l,\beta,q}$ is a matrix whose rows are vectors
    $\tau^{(\beta,\xi_1)},\ldots,
    \tau^{(\beta,\xi_{r(l,\beta,q)})}.$
    Then elements of only one row of the $l$-th array are used
    in the $q$-th access of the operation $S_{\beta}$ for fixed values
    of indices of $d$ outer loops.
\end{theorem}

That is, to obtain space localization is to get $r(l,\beta,q)$
linear independent vectors $\tau^{(\beta,\xi)}$ that satisfy
condition~(\ref{splcon1}); values $\xi$ are intended to be as
small as possible. Condition~(\ref{splcon1}) can be written in the
vector-matrix form\ $
     \widetilde{\tau}^{({\xi})} D_{l,\beta ,q}=0.
$

Thus conditions of space localization can be reduced to vectors
$z_{l,\beta,q}$ minimization (zeroing if it is possible) when the
following conditions are valid
\begin{equation}
    \begin{array}{c}
        \left| \widetilde{\tau}^{({\xi})}D_{l,\beta,q}\right|-
        z_{l,\beta ,q}=0 \enspace.
    \end{array}
\label{space_z}
\end{equation}

\section{Procedure of Affine Transformation of Loop Nests}

Introduce some notation:
$D^{(1)}_{\varphi}$ and $D^{(1)}$ are
sets of matrices $D_{\alpha,\beta}^{\varphi}$ and
$D_{\alpha,\beta}$ accordingly that describe flow-, out- and
anti-dependences;
$D^{(1)}_{\varphi,\ in}$ and $D^{(1)}_{in}$ are sets of matrices
$D_{\alpha,\beta}^{\varphi}$ and $D_{\alpha,\beta}$ accordingly
that describe in-dependences;
$D_F,$ $D_G,$ $D_f$ are sets of matrices $\Delta^F_{l,\beta,q},$
$\Delta^G_{l,\beta,q},$ and vectors $\Delta^f_{l,\beta,q}$
accordingly;
$D^{(1)}_{s}$ is a set of matrices $D_{l,\beta,q};$
$S_\beta^{(1)}=\bbbz^{n_\beta};\ $
$T^{(\beta)}_{1:0}=0^{(n_{\beta})};$
$T^{(\xi)}_{l,\beta,q}$ is a matrix whose rows are vectors
$\tau^{(\beta,d)},\ 1\le d\le\xi,$ satisfying
condition~(\ref{splcon1}); $L^{(\xi)}=\{\, \beta \mid
n-\xi+1=n_\beta-\rrang T^{(\beta)}_{1:{\xi-1}} \,\};$
$\rho(z^\varphi_{\alpha,\beta},z_{\alpha,\beta},z^F_{l,\beta,q},z^G_{l,\beta,q},z^f_{l,\beta,q},z_{l,\beta,q})=
    \msum_{\alpha,\beta}
    \big(\lambda^\varphi_{\alpha,\beta} z^\varphi_{\alpha,\beta}+
    \lambda_{\alpha,\beta} z_{\alpha,\beta}\big)
    +\msum_{l,\beta,q}
    \big(\lambda^F_{l,\beta,q} z^F_{l,\beta,q}+
    \lambda^G_{l,\beta,q} z^G_{l,\beta,q}+
    \lambda^f_{l,\beta,q} z^f_{l,\beta,q}\big)
    +\msum_{l,\beta,q}
    \lambda_{l,\beta,q} z_{l,\beta,q}.$
The sum $\sum\limits_{\alpha,\beta}$ is over all $\alpha, \beta$
such that $D_{\alpha,\beta}\in D^{(\xi)}\cup D^{(\xi)}_{in},$
$D_{\alpha,\beta}^{\varphi}\in D^{(\xi)}_{\varphi}\cup
D^{(\xi)}_{\varphi,in},$ and the sum $\sum\limits_{l, \beta, q}$
is over all $l, \beta, q$ such that $D_{l, \beta ,q}\in
D^{(\xi)}_s,$ $\Delta^F_{l,\beta,q}\in D_F,$
$\Delta^G_{l,\beta,q}\in D_G,$ $\Delta^f_{l,\beta,q}\in D_f;$
$\lambda_{\alpha,\beta}^{\varphi},\ $ $\lambda_{\alpha,\beta},\ $
$\lambda^F_{l,\beta,q},\ $ $\lambda^G_{l,\beta,q},\ $
$\lambda^f_{l,\beta,q},\ $ $\lambda_{l, \beta ,q}$ are weights.
The sets $D^{(\xi)}_{\varphi},$ $D^{(\xi)},$ and
$D^{(\xi)}_{\varphi,in},$ $D^{(\xi)}_{in}$ consist of matrices
$D^\varphi_{\alpha,\beta},$ $D_{\alpha,\beta},$ the sets
$D^{(\xi)}_s$ consist of matrices $D_{l,\beta,q}$ from
conditions~(\ref{depend_z}), (\ref{depend_z_in}), and
(\ref{space_z}) whom the vector $\widetilde\tau^{(\xi)},\ \xi>1,$
is to satisfy.

Coordinates of the weights correspond to columns of the matrices
$D_{\alpha,\beta}^{\varphi},\ $ $D_{\alpha,\beta},\ $
$\Delta^F_{l,\beta,q},\ $ $\Delta^G_{l,\beta,q},\ $ $D_{l, \beta
,q}$ and vectors $\Delta^f_{l,\beta,q}.$ Suppose a column of a
matrix or a vector is found $\gamma$ times; then the larger
$\gamma$ the greater role of this column or vector in minimization
of the number of communications between processors and in
improvement of locality. Thus, the value of appropriate coordinate
is to be larger. The weights can also express the preference for
the choice of operation and data allocation. Suppose it is
desirable that there is no exchange of elements of some array
$a_{l_0}$, then the weights $z^F_{l_0,\beta,q},$
$z^G_{l_0,\beta,q},$ $z^f_{l_0,\beta,q}$ are to be larger then the
others.

To find the vectors $\widetilde{\tau}^{({\xi})}$ it is necessary
to minimize values of the variables $z^{\varphi}_{\alpha,\beta},$
$z_{\alpha,\beta},$ $z^F_{l,\beta,q},$ $z^G_{l,\beta,q},$
$z^f_{l,\beta,q},$ $z_{l,\beta,q}.$ Thus, to find these vectors is
to solve the following optimization problem. Choose a vector
$s_\beta^{(\xi)}\in S_\beta^{(\xi)},\ $ $\beta\in L^{(\xi)},$ and
minimize the value of the function $\rho,$ the following condition
being valid: condition~(\ref{trang}) for $\beta\in L^{(\xi)},$
conditions~(\ref{Joff_z}) if $\xi\le r,$ and
conditions~(\ref{depend_z}), (\ref{depend_z_in}), (\ref{space_z}).

The following procedure summarizes the previous investigations.
The aim of the procedure is to find a multi-dimensional scheduling
and data allocation satisfying the condition of communication-free
allocation and the condition of space and time localization. The
procedure is recursive and consists of $n$ recursions. The $\xi$th
recursion results in getting a vector $\widetilde{\tau}^{(\xi)}.$

\smallskip {\bf Procedure} (finding scheduling and allocation
functions): Put $\xi=1.$\\
{\it Step 1.} Choose a vector $s_\beta^{(\xi)}\in
S_\beta^{(\xi)},\ $ $\beta\in L^{(\xi)}.$ Find a vector
$\widetilde{\tau}^{(\xi)}$ by solving the optimization problem\\
$
 \min\Big\{
 \rho(z^\varphi_{\alpha,\beta},z_{\alpha,\beta},
 z^F_{l,\beta,q},z^G_{l,\beta,q},z^f_{l,\beta,q},z_{l,\beta,q})
 \;\Big|\quad$
 condition (\ref{trang}),\quad $\beta\in L^{(\xi)},$
\\
\hphantom{$\min\Big\{$}
 condition (\ref{depend_z}),\quad
$
 D_{\alpha,\beta}^{\varphi}\in D^{(\xi)}_{\varphi},\
 D_{\alpha,\beta}\in D^{(\xi)},
$\\
\hphantom{$\min\Big\{$}
 condition (\ref{depend_z_in}),\quad
$
 D_{\alpha,\beta}^{\varphi}\in D^{(\xi)}_{\varphi,\ in},
 D_{\alpha,\beta}\in D^{(\xi)}_{in},
$\\
\hphantom{$\min\Big\{$}
 condition (\ref{Joff_z}),\quad
$
 \Delta^G_{l,\beta,q}\in D_G,\
 \Delta^f_{l,\beta,q}\in D_f,\
 \Delta^F_{l,\beta,q}\in D_F,\
 \xi\le r,
$\\
\hphantom{$\min\Big\{$}
 condition (\ref{space_z}),\quad
$
 D_{l, \beta, q}\in D^{(\xi)}_s \Big\}.
$\\
{\it Step 2.} If $\xi< r+1$ then define sets:
$D^{(\xi+1)}_{\varphi}=D^{(\xi)}_{\varphi},$\ \
$D^{(\xi+1)}=D^{(\xi)}.$\\
If $\xi\ge r+1$ then define sets:\\
$D^{(\xi+1)}_{\varphi}=D^{(\xi)}_{\varphi}\backslash\, \{\,
D_{\alpha,\beta}^{\varphi} \mid z_{\alpha,\beta}^{\varphi}>0
\,\},$\ \ $D^{(\xi+1)}=D^{(\xi)}\backslash\, \{\, D_{\alpha,\beta}
\mid z_{\alpha,\beta}^{\varphi}>0 \,\}.$\\
{\it Step 3.} Define sets:\\
$D^{(\xi+1)}_{\varphi,\
in}=D^{(\xi)}_{\varphi,\ in}\backslash\, \{\,
D_{\alpha,\beta}^{\varphi} \mid z_{\alpha,\beta}^{\varphi}>0
\,\},$\ \
$D^{(\xi+1)}_{in}=D^{(\xi)}_{in}\backslash\, \{\,
D_{\alpha,\beta} \mid z_{\alpha,\beta}^{\varphi}>0 \,\},$\\
$D^{(\xi+1)}_{s}=\{\, D_{l, \beta ,q}\in D_s^{(1)} \mid \rrang
T_{l,\beta,q}^{(\xi)}<r(l,\beta,q) \,\}.$\\
{\it Step 4.} Define a set $L^{(\xi+1)}=\{\, \beta \mid n-\xi
=n_\beta - \rrang T^{(\beta)}_{1:\xi} \,\}.$\\
{\it Step 5.} If $\xi=n$ then go out the procedure else increase
$\xi$ by 1 and go to step~1.

\section{Data Exchange Sequence}

Suppose for some fixed parameters $l,\beta,q,\xi$ the conditions
of communication-free allocation are not valid (i.e., even one of
variables $z^F_{l,\beta,q},$ $z^G_{l,\beta,q},$ $z^f_{l,\beta,q}$
is not equal to zero at some recursion of the procedure). Then it
is necessary to pass elements of array $a_l$ for using them for
the $q$-th input of elements of array $a_l$ into instruction
$S_\beta.$

By $P(z_1,\ldots, z_r)$ denote a processor allocated at the point
$(z_1,\ldots, z_r)$ of virtual processors space. According to the
functions $\overline d^{(l)}$ and $\overline t^{(\beta)},$ the
array elements $a_l(\overline F_{l,\beta,q}(J))$ are stored in the
local memory of the processors
$P(d^{(l)}_1(\overline{F}_{l,\beta,q}(J)),\linebreak\ldots,
d^{(l)}_r(\overline{F}_{l,\beta,q}(J)))$ and they are used in the
processors $P(t^{(\beta)}_1(J),\ldots,t^{(\beta)}_r(J))$ at the
iterations $(t^{(\beta)}_{r+1}(J),\ldots,t^{(\beta)}_n(J)).$ In
the general case, point-to-point communications can be organized
between pairs of these processors.)

For the program execution time to be smaller it is desirable to
determine prompt communications such as broadcast, gather,
scatter, reduction, and data translation. Consider for example
broadcast.

Let $F$ be an element of the set $W_l$. Denote by
$V_{l,\beta,q}^{(F)} = \{\; J\in V_\beta \mid \overline
F_{l,\beta,q}(J)=F \;\} $ the set of such iterations of the
initial loop nest that the array element $a_l(F)$ is used at them
for the $q$-th input of elements of array $a_l$ into instruction
$S_\beta.$ The set $V_{l,\beta,q}^{(F)}$ is called non-degenerate
if ${\rm dim}(\ker F_{l,\beta,q})\neq0$ and there exists a vector
$J_0\in V_{l,\beta,q}^{(F)}$ such that $J_0+u_i\in V_\beta,$ where
$u_i$ is any base vector of the intersection $\ker F_{l,\beta,q}$
and $\bbbz^{n_\beta}.$

Let $u^{(1)}_{l,\beta,q},\ldots,
u^{(\zeta(l,\beta,q))}_{l,\beta,q}$ be a fundamental system of
solutions of a uniform system of equations $F_{l,\beta ,q}x=0.$

\begin{theorem}
    Suppose the set $V_{l,\beta,q}^{(F)}$ is non-degenerate;
    the function $\overline{F}_{l,\beta,q}$ occurs in the right part of
    the instruction $S_\beta;$ conditions
    $$
        \tau^{(\beta,\xi)} u^{(\zeta)}_{l,\beta,q}=0,\quad
        r+1\le\xi\le n,\  1\le\zeta\le\zeta(l,\beta,q)\enspace,
    $$
    and one of the following conditions are valid:\\
    a) the elements of array $a_l$ occur only in the right parts of
       the instructions,\\
    b) constraints
       $$
            \Phi_{\alpha,\beta} u^{(\zeta)}_{l,\beta,q}=0,\quad
            1\le\zeta\le\zeta(l,\beta,q)\enspace,
       $$
       are valid for the flow-dependence produced by the $q$-th input
       of elements of array $a_l$ into instruction $S_\beta.$\\
    Then to pass the data $a_l(F)$ it is possible to arrange
    broadcast from the processor
    $P(d^{(l)}_1(\overline{F}_{l,\beta,q}(J)),\ldots,
    d^{(l)}_r(\overline{F}_{l,\beta,q}(J)))$ to the processors
    $P(t^{(\beta)}_1(J),\ldots,t^{(\beta)}_r(J))$ at the iteration
    $(t^{(\beta)}_{r+1}(J),\ldots,t^{(\beta)}_n(J)),$ $J\in
    V_{l,\beta,q}^{(F)}.$
\end{theorem}

\section{Conclusion}
In this paper, we propose a method of mapping algorithms for
parallel execution onto distributed memory parallel computers. The
method provides with determination of operation and data
allocation over processors, an execution sequence of operations,
and data exchange necessary for the program execution. The aim is
to minimize a number of communications, to improve locality of an
algorithm, and to determine the possibility of broadcasts.

Note some advantages of the method suggested:\\
-- an initial algorithm is represented by affine loop nests of an
arbitrary nesting structure;\\
-- the suggested conditions can be simply obtained
from a source algorithm;\\
-- the conditions do not depend on the definite values of outer
variables; the obtained functions depend on outer variables
parametrically;\\
-- the method can be automated.

The method was applied for mapping algorithms for matrix
transformations onto distributed memory parallel computers. These
algorithms was implemented on the supercomputer SKIF (it is
located at NAS of Belarus, Minsk).


\end{document}